\documentstyle[wstwocl,epsfig]{article}
\begin{document}

\bibliographystyle{unsrt}    

\newcommand{\st}{\scriptstyle}
\newcommand{\sst}{\scriptscriptstyle}
\newcommand{\mco}{\multicolumn}
\newcommand{\epp}{\epsilon^{\prime}}
\newcommand{\vep}{\varepsilon}
\newcommand{\ra}{\rightarrow}
\newcommand{\ppg}{\pi^+\pi^-\gamma}
\newcommand{\vp}{{\bf p}}
\newcommand{\ko}{K^0}
\newcommand{\kb}{\bar{K^0}}
\newcommand{\al}{\alpha}
\newcommand{\ab}{\bar{\alpha}}
\def\be{\begin{equation}}
\def\ee{\end{equation}}
\def\bea{\begin{eqnarray}}
\def\eea{\end{eqnarray}}
\def\CPbar{\hbox{{\rm CP}\hskip-1.80em{/}}}

\def\ap#1#2#3   {{\em Ann. Phys. (NY)} {\bf#1} (#2) #3.}
\def\apj#1#2#3  {{\em Astrophys. J.} {\bf#1} (#2) #3.}
\def\apjl#1#2#3 {{\em Astrophys. J. Lett.} {\bf#1} (#2) #3.}
\def\app#1#2#3  {{\em Acta. Phys. Pol.} {\bf#1} (#2) #3.}
\def\ar#1#2#3   {{\em Ann. Rev. Nucl. Part. Sci.} {\bf#1} (#2) #3.}
\def\cpc#1#2#3  {{\em Computer Phys. Comm.} {\bf#1} (#2) #3.}
\def\err#1#2#3  {{\it Erratum} {\bf#1} (#2) #3.}
\def\ib#1#2#3   {{\it ibid.} {\bf#1} (#2) #3.}
\def\jmp#1#2#3  {{\em J. Math. Phys.} {\bf#1} (#2) #3.}
\def\ijmp#1#2#3 {{\em Int. J. Mod. Phys.} {\bf#1} (#2) #3.}
\def\jetp#1#2#3 {{\em JETP Lett.} {\bf#1} (#2) #3.}
\def\jpg#1#2#3  {{\em J. Phys. G.} {\bf#1} (#2) #3.}
\def\mpl#1#2#3  {{\em Mod. Phys. Lett.} {\bf#1} (#2) #3.}
\def\nat#1#2#3  {{\em Nature (London)} {\bf#1} (#2) #3.}
\def\nc#1#2#3   {{\em Nuovo Cim.} {\bf#1} (#2) #3.}
\def\nim#1#2#3  {{\em Nucl. Instr. Meth.} {\bf#1} (#2) #3.}
\def\np#1#2#3   {{\em Nucl. Phys.} {\bf#1} (#2) #3.}
\def\pcps#1#2#3 {{\em Proc. Cam. Phil. Soc.} {\bf#1} (#2) #3.}
\def\pl#1#2#3   {{\em Phys. Lett.} {\bf#1} (#2) #3.}
\def\prep#1#2#3 {{\em Phys. Rep.} {\bf#1} (#2) #3.}
\def\prev#1#2#3 {{\em Phys. Rev.} {\bf#1} (#2) #3.}
\def\prl#1#2#3  {{\em Phys. Rev. Lett.} {\bf#1} (#2) #3.}
\def\prs#1#2#3  {{\em Proc. Roy. Soc.} {\bf#1} (#2) #3.}
\def\ptp#1#2#3  {{\em Prog. Th. Phys.} {\bf#1} (#2) #3.}
\def\ps#1#2#3   {{\em Physica Scripta} {\bf#1} (#2) #3.}
\def\rmp#1#2#3  {{\em Rev. Mod. Phys.} {\bf#1} (#2) #3.}
\def\rpp#1#2#3  {{\em Rep. Prog. Phys.} {\bf#1} (#2) #3.}
\def\sjnp#1#2#3 {{\em Sov. J. Nucl. Phys.} {\bf#1} (#2) #3.}
\def\spj#1#2#3  {{\em Sov. Phys. JEPT} {\bf#1} (#2) #3.}
\def\spu#1#2#3  {{\em Sov. Phys.-Usp.} {\bf#1} (#2) #3.}
\def\zp#1#2#3   {{\em Zeit. Phys.} {\bf#1} (#2) #3.}

\setcounter{secnumdepth}{2} 

\title{SUPERSYMMETRIC SCATTERING IN TWO DIMENSIONS\footnotemark\ }

\firstauthors{M. Moriconi}

\firstaddress{Joseph Henry Laboratories, Princeton University, Princeton,
NJ 08544, USA}

\secondauthors{K. Schoutens}

\secondaddress{Institute for Theoretical Physics, University of
Amsterdam, Valckenierstraat 65, 1018 XE Amsterdam, The Netherlands}

\twocolumn[\maketitle\abstracts{We briefly review results on
two-dimensional supersymmetric quantum field theories that
exhibit factorizable particle scattering. Our particular
focus is on a series of $N\!=\!1$ supersymmetric theories,
for which exact $S$-matrices have been obtained. A Thermodynamic
Bethe Ansatz (TBA) analysis for these theories has confirmed the
validity of the proposed $S$-matrices and has pointed at an
interesting `folding' relation with a series of $N\!=\!2$ supersymmetric
theories.}]

\section{Introduction}

\footnotetext{talk delivered at the HEP95 Conference of
              the EPS, Brussels, July/August 1995}
To many of the practitioners of experimental or theoretical
High Energy Physics, the subject of integrable Quantum Field Theories
(QFT) in $1\!+\!1$ (Minkowski) space-time dimensions may seem like a
rather remote corner of their field. Real particles (and people, for
that matter) live in four dimensions and the field theories that are
of direct relevance for Particle Physics are certainly not integrable.
Thus, work on QFT in $1\!+\!1$ dimensions may seem to be a rather pointless
exercise.

We would like to use the space allotted to us here to argue to the
contrary, and to illustrate our point by citing some recent results
\cite{kjs,ms1} on supersymmetric particle scattering in $d\!=\!1\!+\!1$
dimensions.

One fact of life in four dimensions is that some of the QFT's that we
need to do physics, in particular Quantum Chromodynamics (QCD), cannot
satisfactorily be analyzed on the basis of perturbation theory alone,
and that some of the essential physics in those theories is of
non-perturbative nature. We do know of certain lower dimensional QFT's
that exhibit some of those same phenomena, and that are at the same time
integrable, i.e. exactly solvable. Such lower dimensional theories
offer a testing ground where the results of perturbation theory
together with non-perturbative techniques can be tested against
exact solutions. As an example we may cite the recent studies of
the two dimensional version of QCD (QCD$_2$), which have in
particular focused on the possibility of a string theoretical
formulation of some of the strong coupling effects in the QCD$_4$.

More direct applications of $d\!=\!1\!+\!1$ QFT become possible
as soon as the essential physics takes place in the radial
direction of three-dimensional space. The radial coordinate $r$
lives on a half-line, and taking  only radial degrees of freedom into
account places us directly in the context of $d\!=\!1\!+\!1$ QFT.
Note that such applications require an understanding of
what happens at the boundary of the half-line, i.e. at the
origin of three-dimensional space. The answer to this question may become
very interesting if we assume the presence of a non-trivial
object, such as a magnetic monopole or a black hole, at the
origin. Phenomena such as the Callan-Rubakov effect (the catalysis
of baryon decay through magnetic monopoles) and the Hawking
effect (the quantum radiation of black holes) may thus be analyzed
using QFT's that are essentially one-dimensional.
In general, the treatment of non-trivial boundary interactions
requires some care, as the bulk QFT may be affected in a rather
drastic way. In recent years, techniques to analyze boundary
theories (both conformal and massive) have been developed
and notions such as integrability have successfully been extended
to the situation with boundary.

Yet another way in which 2d QFT is of relevance
is of course in string theory,
which is largely formulated as a $d=2$ QFT on a dynamical
(world-sheet) surface.

Last but not least we would like to stress the fact that the
subject of $d\!=\!1\!+\!1$ (integrable) QFT is one of those areas
where theoretical physicists and mathematicians with entirely
different backgrounds come together and share ideas.
Two-dimensional field theories are of great interest for
Statistical Mechanics and Condensed Matter Theory. For example,
the aforementioned boundary theories have found beautiful
applications in the analysis of the (multi-channel) Kondo problem
and of problems involving the tunneling of edge currents in
Fractional Quantum Hall devices. Thus, as it has often
happened in the history of Physics, theorists working on
entirely different physical problems are sharing the same
tools and formalisms. Once recognized, this insight (which tends
to be ignored by many of our funding and publishing agencies)
can be used to the benefit of all involved.

\section{Supersymmetry in two dimensions}

Supersymmetric extensions of $1\!+\!1$ dimensional QFT's are
entirely natural from the point of view of supersymmetric
particle theories in four dimensions and superstrings.
Superstrings may be viewed as world sheet field theories with
superconformal invariance, i.e. with symmetries that form
a supersymmetric extension of the conformal (Virasoro) algebra.
In general, superconformal field theories may be perturbed
by relevant perturbations that respect the supersymmetry,
and this then leads to massive supersymmetric theories.
For well-chosen perturbations the resulting massive theory
may be integrable, by which we mean that it possesses an
infinite number of non-trivial integrals of motion beyond
energy and momentum.

Around 1990, many examples of integrable supersymmetric
massive particle theories in $d\!=\!1\!+\!1$ dimensions have been
studied, the majority possessing either $N\!=\!1$ or $N\!=\!2$
supersymmetry. These theories are typically characterized
by an explicit lagrangian or, as outlined above, as a
perturbation of a superconformal field theory. One of the
great challenges is to determine the exact many-particle
scattering amplitudes, i.e. the $S$-matrix. For this problem
the property of integrability turns out to be of great help,
since that implies the {\em factorizability} of the $S$-matrix,
meaning that particle production is entirely forbidden and
that all $N$-particle amplitudes are simply obtained as products
of $2$-particle scattering matrices. The consistency of such a
decomposition requires a special property of the $2$-particle
$S$-matrices, the so-called Yang-Baxter equation.
Additional consistency conditions arise from the compatibility
of the higher integrals of motion with the bound-state structure
of the $S$-matrix. Together these conditions are so restrictive
that one may approach the problem of finding an exact $S$-matrix
by what is called a `bootstrap approach': one uses the
consistency conditions to construct a set of `minimal' consistent
$S$-matrices and then tries to match those with the theories at hand.

One way to test the conjectured identification of an exact $S$-matrix
is through the so-called Thermodynamic Bethe Ansatz (TBA), which is
a procedure that allows one to derive thermodynamical properties
of field theories specified by a (factorizable) $S$-matrix.
Some of these thermodynamic quantities, such as the central
charge and the scaling dimensions of the ultraviolet limit of
the theory, are known from the start and can thus be used as
a non-trivial consistency check on a proposed $S$-matrix.

In a 1990 paper \cite{kjs}, one of us considered the simplest type
of supersymmetry in $1\!+\!1$ dimensions, which is characterized by a
$1\!+\!1$ dimensional super-Poincar\'e algebra without central
charges
\be
Q^2 = P \ , \quad
\overline{Q}^2 = \overline{P}\ , \quad
\{ Q, \overline{Q} \} = 0 \ .
\ee
The paper \cite{kjs} proposed a set of exact scattering matrices of
the form
\be
   S = S^{[ij]}_{B} \, S^{[ij]}_{BF}, \quad i,j=1,2,\ldots,n \, ,
\label{smat}
\ee
where $S^{[ij]}_{B}$ is a diagonal $S$-matrix for a collection
of bosonic particles $\{b_1,b_2,\ldots,b_n\}$, of masses
$m_1,m_2,\ldots,m_n$, and $S^{[ij]}_{BF}$ is a universal `Bose-Fermi'
$S$-matrix that describes the mixing of these bosonic particles
with the corresponding fermions $\{f_1,f_2,\ldots,f_n\}$.
In words, the formula (\ref{smat}) describes a theory that is a
supersymmetrization of a bosonic theory with diagonal scattering.
Notice that the full $S$-matrix (\ref{smat}) is non-diagonal.

The paper [1] showed that supersymmetry and factorizability
alone determine the form of the Bose-Fermi $S$-matrix $S^{[ij]}_{BF}$
up to one free constant $\alpha$. Furthermore, it derived a
condition for the consistency of the full $S$-matrix, stating that
as soon as the particle $b_k$ may occur (according to the bosonic
factor $S_B^{[ij]}$ of the $S$-matrix) as a bound state of particles
$b_i$ and $b_j$, the following relation should hold
\be
\alpha = - {(2m_i^2m_j^2+2m_i^2m_k^2+2m_j^2m_k^2
                  - m_i^4 - m_j^4 - m_k^4)^{1 \over 2} \over
            2 m_i m_j m_k} \ .
\label{crit}
\ee
Note that this condition is extremely restrictive, since
one single free parameter $\alpha$ should be adjusted to
accommodate a large number of non-zero three-point couplings.
The three-point couplings $f_{ijk}$ of the bosonic
theory branch into couplings $f_{f_if_jb_k}$, $f_{f_ib_jf_k}$,
$f_{b_if_jf_k}$ and $f_{b_ib_jb_k}$, whose ratio's depend
only on the masses of the particles involved, according to
\be
 \frac{f_{f_if_jb_k}}{f_{b_ib_jb_k}} =
 \left( \frac{m_i+m_j-m_k}{m_i+m_j+m_k} \right)^{1 \over 2} \ .
\ee

In ref.~[1] it was proposed that a specific series of perturbed
$N\!=\!1$ superconformal field theories lead to a scattering theory
that is precisely of the above type. The superconformal
field theories are the non-unitary minimal superconformal
models labeled as $M(4n+4,2)$, of central charge
\be
c_n = - {3n(4n+3) \over 2(n+1)} \ ,
\label{cc}
\ee
and the perturbing operator is the bottom component of the
Neveu-Schwarz superfield labeled as $\phi_{(1,3)}$.
The resulting massive theories contain $n$ $N\!=\!1$
multiplets $(b_i,f_i)$ of mass
\be
  m_i = \frac{\sin(i\beta\pi)}{\sin(\beta\pi)} \ , \quad
  \beta = {1 \over 2n+1} \ ,
\label{masses}
\ee
and the parameter $\alpha$ takes the value
\be
\alpha = \alpha_n = - sin(\beta \pi) \ .
\ee
The bosonic factor $S_B$ of the $S$-matrix is identical
to the $S$-matrix first described in [3], and
the corresponding bound state structure is precisely such that
the criterion (\ref{crit}) is satisfied for all non-vanishing
oouplings.

\section{TBA and the relation with $N\!=\!2$}

The Thermodynamic Bethe Ansatz (TBA) for the series of $N\!=\!1$
supersymmetric $S$-matrices has recently been completed \cite{ms1},
see also [4]. At the technical level, this has been somewhat
of a tour de force due to the fact that the full $N\!=\!1$ $S$-matrix
is non-diagonal. This implies that the fundamental TBA equation for the
rapidities of $L$ particles that live on a circle contains
a matrix of dimension $2^L \times 2^L$, and in order to complete
the TBA the eigenvalues of this matrix have to be determined.
After re-interpreting the 2-particle $S$-matrices as Boltzmann
weights, one may recognize the transfermatrix as coming from an
eight-vertex model. For the explicit evaluation of these eigenvalues
we have, following [5,4] relied on the fact that the
Boltzmann weights satisfy what is called a `free fermion
condition', which allows one to avoid the generalized Bethe
Ansatz that is needed for the general eight-vertex model.

The results of the TBA analysis have confirmed, among other things,
the value (\ref{cc}) of the ultraviolet central charge and,
thereby, the validity of the proposed $S$-matrices for the
perturbed superconformal theories. We refer to our paper \cite{ms1}
for the details of this analysis.

With the TBA for the $N\!=\!1$ theory completed, we are in a position
to compare a number of theories \cite{km,fi,kjs} that all have the
mass spectrum given by (\ref{masses}) and that have essentially the
same fusion rules,
\begin{enumerate}
\item
$2n$ bosonic particles $b_i$, $\bar{b}_i$, $i=1,2,\ldots,n$,
perturbation of CFT of central charge $c={4n \over 2n+3}$,
\item
$n$ bosonic particles $b_i$, $i=1,2,\ldots,n$,
perturbation of CFT of central charge $c=-{2n(6n+5) \over 2n+3}$,
\item
$4n$ particles $b_i$, $\bar{b}_i$, $f_i$, $\bar{f}_i$, $i=1,2,\ldots,n$,
$N\!=\!2$ supersymmetric perturbation of $N\!=\!2$ CFT of central charge
$c={3n \over n+1}$,
\item
$2n$ particles $b_i$, $f_i$, $i=1,2,\ldots,n$,
$N\!=\!1$ supersymmetric perturbation of $N\!=\!1$ CFT of central charge
$c=-{3n(4n+3) \over 2(n+1)}$.
\end{enumerate}
The mass spectrum and fusion rules that these theories have in
common are clearly linked to the Lie algebra $A_{2n}$;
the first two theories in this list are usually denoted as
the $A_{2n}^{(1)}$ and $A_{2n}^{(2)}$ scattering theories,
respectively.

Let us now briefly discuss the relations among the theories
in this list. Clearly, theory 3 is an $N\!=\!2$ supersymmetrization
of theory 1 and, similarly, theory 4 is an $N\!=\!1$ supersymmetrization
of theory 2. In addition, theories 2 and 4 are obtained from
1 and 3 by eliminating half of the particles from the theory.
Between the theories 1 and 2 there is a simple (multiplicative)
relation \cite{alz,km} at the level of the $S$-matrices,
which leads to an additive relation at the level of the TBA equations.
This relation has been called `folding in half'.
For the TBA systems of the $N\!=\!2$ and $N\!=\!1$ supersymmetric theories
(theories 3 and 4) a similar `folding' relation was conjectured
by E. Melzer \cite{mel}. Our explicit computation of the $N\!=\!1$
TBA equations have confirmed this conjecture and established the
folding relation. It is expected that this folding can be traced
back to a folding relation at the level of the $N\!=\!2$ and $N\!=\!1$
$S$-matrices, but this has not been worked out. [Writing folding
relations at the $S$-matrix level is not straightforward since the
Yang-Baxter equation, which is automatic in the bosonic case,
poses severe restrictions in the supersymmetric version.]

Before closing, we would like to stress that the entire
structure discussed here can be extended to $N\!=\!1$ supersymmetric
boundary theories. In that case the $S$-matrix data are supplemented
by a boundary reflection matrix $R$, which is subject to a
boundary Yang-Baxter equation. A publication on these matters
is in preparation \cite{ms2}.

\setcounter{secnumdepth}{0} 

\section{Acknowledgments}
It is a pleasure to thank Ezer Melzer, Roland Koberle, Paul Fendley
and Omar Foda for discussions in various stages of this work.
M.M. was supported by a fellowship from CNPq (Brazil) and
K.S. was supported in part by the foundation FOM (the Netherlands).

\end{document}